\documentclass[aps,prx,twocolumn,superscriptaddress,showpacs,longbibliography,nofootinbib]{revtex4-1}
\usepackage{amsmath}
\usepackage{graphicx}
\usepackage{amsfonts}
\usepackage{verbatim}
\usepackage{amssymb}
\usepackage{color}
\usepackage{dsfont}
\usepackage{amsmath} 
\usepackage{hyperref}
\usepackage{physics}
\usepackage{soul}
\hypersetup{colorlinks = true, urlcolor = blue, linkcolor = blue, citecolor = blue}
\usepackage[normalem]{ulem}

\newcommand{\bpm}{\begin{pmatrix}}
\newcommand{\epm}{\end{pmatrix}}

\newcommand{\nn}{\nonumber \\} 
\newcommand{\tp}{ ^{\intercal} }
\newcommand{\dg}{^{\dagger}}
\newcommand{\hc}{\rm{H.c.}}

\begin{document}

\title{Fusion protocol for Majorana modes in coupled quantum dots}
\author{Chun-Xiao Liu$^{*,\dagger}$}
\affiliation{Qutech and Kavli Institute of Nanoscience, Delft University of Technology, Delft 2600 GA, The Netherlands}

\author{Haining Pan$^*$}
\affiliation{Condensed Matter Theory Center and Joint Quantum Institute, Department of Physics, University of Maryland, College Park, Maryland 20742-4111, USA}
\affiliation{Department of Physics, Cornell University, Ithaca, NY, USA}

\author{F. Setiawan$^\P$}
\affiliation{Pritzker School of Molecular Engineering, University of Chicago, 5640 South Ellis Avenue, Chicago, Illinois 60637, USA}
\affiliation{Riverlane Research Inc., Cambridge, 02142, USA}

\author{Michael Wimmer}
\affiliation{Qutech and Kavli Institute of Nanoscience, Delft University of Technology, Delft 2600 GA, The Netherlands}

\author{Jay D. Sau}
\affiliation{Condensed Matter Theory Center and Joint Quantum Institute, Department of Physics, University of Maryland, College Park, Maryland 20742-4111, USA}
\date{\today}

\begin{abstract}
In a recent breakthrough experiment [Nature 614, 445-450 (2023)], signatures of Majorana zero modes have been observed in tunnel spectroscopy for a minimal Kitaev chain constructed from coupled quantum dots.
However, as Ising anyons, Majoranas' most fundamental property of non-Abelian statistics is yet to be detected. 
Moreover, the minimal Kitaev chain is qualitatively different from topological superconductors in that it supports Majoranas only at a sweet spot.
Therefore it is not obvious whether non-Abelian characteristics such as braiding and fusion can be demonstrated in this platform with a reasonable level of robustness.
In this work, we theoretically propose a protocol for detecting the Majorana fusion rules in an artificial Kitaev chain consisting of four quantum dots.
In contrast with the previous proposals for semiconductor-superconductor hybrid nanowire platforms, here we do not rely on mesoscopic superconducting islands, which are difficult to implement in quantum dot chains.
To show the robustness of the fusion protocol, we discuss the effects of three types of realistic imperfections on the fusion outcomes, e.g. diabatic errors, dephasing errors, and calibration errors.
We also propose a Fermion parity readout scheme using quantum capacitance.
Our work will shed light on future experiments on detecting the non-Abelian properties of Majorana modes in a quantum dot chain.
\end{abstract}

\maketitle
\def\thefootnote{*}\footnotetext{These authors contributed equally to this work}\def\thefootnote{\arabic{footnote}}
\def\thefootnote{$\dagger$ }\footnotetext{{\color{blue}chunxiaoliu62@gmail.com}}\def\thefootnote{\arabic{footnote}}
\def\thefootnote{$\P$}\footnotetext{Present Address: Riverlane Research, Inc., Cambridge, Massachusetts, 20142, USA}

\section{Introduction}
Majorana zero modes are midgap, charge-neutral quasiparticle excitations localized at the endpoints of a topological superconductor~\cite{Alicea2012New, Leijnse2012Introduction, Beenakker2013Search, Stanescu2013Majorana, Jiang2013Non, Elliott2015Colloquium, Sato2016Majorana, Sato2017Topological, Aguado2017Majorana, Lutchyn2018Majorana, Zhang2019Next, Frolov2020Topological}.
They obey non-Abelian statistics, namely, swapping two Majoranas would transform the many-body wavefunction into a new one within the degenerate ground-state manifold, and thereby can be utilized as the building block for error-resilient topological quantum computation~\cite{Nayak2008Non-Abelian, DasSarma2015Majorana}.
In a very recent experiment~\cite{Dvir2023Realization}, following the theoretical proposals~\cite{Sau2012Realizing, Leijnse2012Parity, Fulga2013Adaptive, Liu2022Tunable}, Majoranas were observed in a minimal Kitaev chain constructed from coupled quantum dots. 
In particular, Majoranas emerge only at the sweet spot of the system, i.e., when dot energies are placed at the Fermi level of the superconductor, and the normal and superconducting couplings are made equal in strength.
While such Majoranas do not have the exponential protection against parameter changes expected for an ideal long topological superconducting wire, they still possess topological properties near the sweet spot.

Motivated by such experimental progress, one may hope to demonstrate some of the defining properties of Majoranas as non-Abelian anyons in quantum dot chains. 
Two equally fundamental properties of non-Abelian anyons are (i) non-Abelian exchange statistics which exhibits in braiding experiments, and (ii) nontrivial fusion rules which can be detected in fusion experiments.
Specifically, Majoranas which are Ising anyons obey the fusion rule:
\begin{align}
    \sigma \times \sigma = I + \Psi,
    \label{eq:fusion_rule}
\end{align}
where two Ising anyons ($\sigma$) fuse into either a vacuum ($I$) or a regular fermion ($\Psi$).
In this work, we focus on the fusion rule detection experiment, which in general requires a much simpler device setup than braiding experiments, and hence is a more attainable goal to pursue in the near future. 
In the nanowire setup~\cite{Mourik2012Signatures, Das2012Zero, Deng2012Anomalous,Churchill2013Superconductor, Finck2013Anomalous, Albrecht2016Exponential, Chen2017Experimental, Deng2016Majorana, Nichele2017Scaling, Zhang2021Large, Wang2022Plateau, Aghaee2022InAs}, different approaches have been proposed to demonstrate the Majorana fusion rules~\cite{Aasen2016Milestones, Hell2016Time, Clarke2017Probability, Zhou2022Fusion, Souto2022Fusion}, most of which require a floating superconducting island with finite charging energy for parity-to-charge conversion that is central to manipulation and readout schemes~\cite{Aasen2016Milestones, Hell2016Time, Souto2022Fusion}.
For the coupled-dot platform, however, the superconductor has to be grounded to induce cross Andreev reflection between quantum dots~\cite{ Sau2012Realizing, Leijnse2012Parity, Liu2022Tunable, Tsintzis2022Creating}, making it difficult to implement finite charging energy~\cite{Wang2022Singlet, Dvir2023Realization, Wang2022Triplet}.
Therefore, a new method to manipulate and read out Majoranas in the quantum dot chain is urgently needed.

\begin{figure}[t]
\centering
{\includegraphics[width = \linewidth]{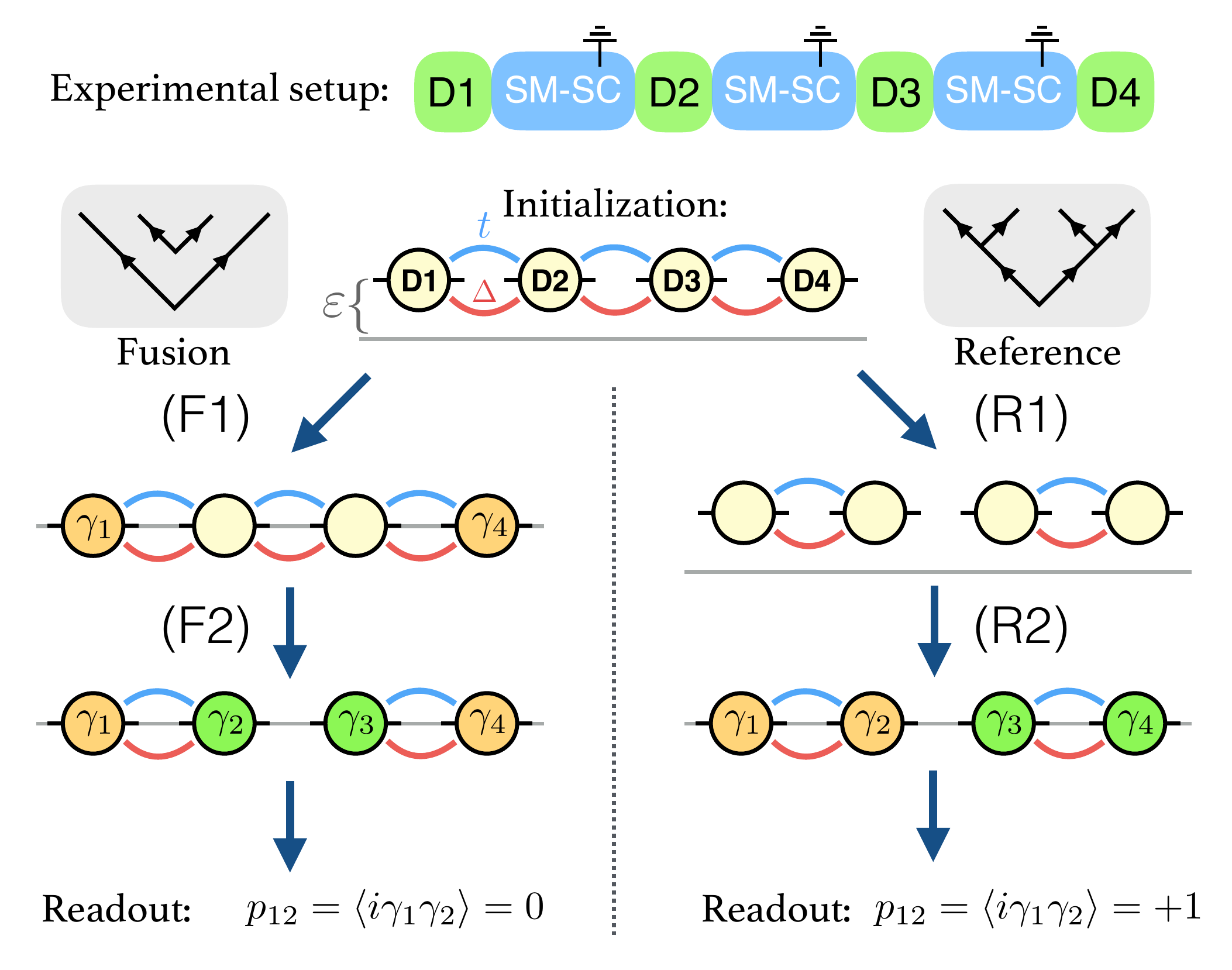}}
\caption{
Upper panel: Schematic of a minimal setup for detecting the Majorana fusion rules.
Initialization: The system is initialized with $\varepsilon_j>0$ and $t_{j,j+1} = \Delta_{j,j+1}>0$ corresponding to a vacuum state with even Fermion parity.
Left panels: Fusion protocol. 
In the first step F1, all four dot energies are tuned from finite to zero, followed by F2 where both the normal and superconducting couplings between dot D2 and D3 are turned off.
It yields a probabilistic Fermion parity readout of $p_{12}=\pm1$ with an average of zero.
Right panels: Reference protocol, which yields a deterministic parity readout $p_{12}=+1$.
Boxes in grey: Fusion diagrams of anyons for the two protocols.
}
\label{fig:fusion_protocol}
\end{figure}

In this work, we propose a minimal setup for detecting the Majorana fusion rules; see Fig.~\ref{fig:fusion_protocol}.
This architecture is also the shortest chain that can support four Majoranas comprising a qubit.
Here, Majoranas are manipulated by changing the effective couplings and dot energies via the electrostatic gates nearby. Meanwhile the readout of the fusion outcome is implemented by quantum capacitance measurement.
All the operations in our protocol are performed in an electrostatic way without the need of floating superconducting islands and without moving Majoranas spatially.
To connect to realistic situations and to show the robustness of our protocol, we analyze the effects of realistic imperfections on the fusion outcomes. Our realistic simulation also allows us to predict the optimal parameter regime for future fusion experiments.

\section{Setup and Hamiltonian}
The setup consists of an array of four dots connected by three hybrid segments in between, as shown in Fig.~\ref{fig:fusion_protocol}.
The effective Hamiltonian of the system is
\begin{align}
\hat{H} = \sum^4_{j=1} \varepsilon_j \hat{n}_j + \sum^3_{j=1} \left( t_{j, j+1} \hat{c}\dg_{j+1} \hat{c}_j + \Delta_{j, j+1} \hat{c}_{j+1} \hat{c}_j + \hc \right),
\label{eq:H_array}
\end{align}
where $\varepsilon_j$ is the dot level energy, $\hat{n}_j$ is the occupancy, $t_{j, j+1}$ and $\Delta_{j,j+1}$ denote the normal and superconducting couplings between adjacent dots, respectively.
In practice, the dots are spin-polarized under a strong magnetic field, and the inter-dot couplings are tunable by changing the properties of Andreev bound states in the hybrid segment~\cite{Liu2022Tunable}.
Here, we assume that all $t_{j, j+1}$ and $\Delta_{j, j+1}$ are real, which is a good approximation for one-dimensional nanowires in the symmetry class BDI~\cite{Tewari2012Topological}.
Under these assumptions, when the system is tuned into its sweet spot, i.e., $\varepsilon_j=0$ and $t_{j, j+1}=\Delta_{j,j+1} = \Delta_0>0$~\cite{Kitaev2001Unpaired} , a pair of Majorana zero modes will emerge and be localized completely at the outermost quantum dots.

\section{Fusion rule protocol}
We first outline our protocols for fusion rule detection in the ideal case, i.e., the system is subject to no noise, and all operations are performed with perfect precision in the adiabatic limit.
The key idea behind testing fusion is to measure a different pairing of Majoranas from the one which was initialized~\cite{Aasen2016Milestones}.
Our system is initialized with $t_{j, j+1} = \Delta_{j, j+1} =\Delta_0 >0$ and $\varepsilon_j>0$, which is a topologically trivial phase corresponding to a vacuum with even Fermion parity.
The final measurement is on the Fermion parity $p_{12}$ at one end of the chain, after the system is ``cut'' into two halves, with each half being tuned to the topological phase.
Specifically, we create the outermost Majoranas $\gamma_{1,4}$ (F1) first, by driving the whole array into the sweet spot where all the dot energies are tuned from finite to zero.
Since the total parity of the initial state was even, i.e., $p_{14}=+1$, the pair of $\gamma_{1,4}$ is initialized to be in the state $| +_{14} \rangle$.
We then cut the middle of the chain, i.e., $t_{23}=\Delta_{23}\to 0$, to reach the measurement step (F2 in Fig.~\ref{fig:fusion_protocol}).
This nucleates the other pair of Majoranas $\gamma_{2,3}$, which is again constrained to be an even state to conserve total parity. 
The resulting final state is
\begin{align}
    | \psi_f \rangle_{F} = | +_{14}, +_{23} \rangle = \left( | +_{12}, +_{34} \rangle + | -_{12}, -_{34} \rangle  \right) / \sqrt{2},
\label{eq:Fusion}
\end{align}
where the second equality is obtained by a basis change related to the $F$ symbols for Ising anyons~\cite{Aasen2016Milestones}.
Equation~\eqref{eq:Fusion} shows that the measurement of the end parity $p_{12}$ yields an 
indeterministic result where $p_{12}=+ 1$ or $-1$ has equal probability. This result can be viewed as evidence for a successful test of the fusion protocol~\cite{Aasen2016Milestones}, even though the non-deterministic result is not uniquely associated with Majoranas~\cite{Clarke2017Probability}.

The fusion protocol test (F in Fig.~\ref{fig:fusion_protocol}) should be contrasted with a reference protocol 
(R in Fig.~\ref{fig:fusion_protocol}) where the Majoranas $\gamma_{1,2,3,4}$ are initialized in the same basis that they are measured in. 
As a result, it gives a deterministic result 
where the measured parity $p_{12}=+1$ and the final state is 
\begin{align}\label{eq:Reference}
     | \psi_f \rangle_R = | +_{12}, +_{34} \rangle.
\end{align}
This reference protocol serves as a baseline since it differs from the fusion protocol $F$ only in whether the chain is cut before or after the dot energies are tuned to zero.

\begin{figure}
\centering
{\includegraphics[width = \linewidth]{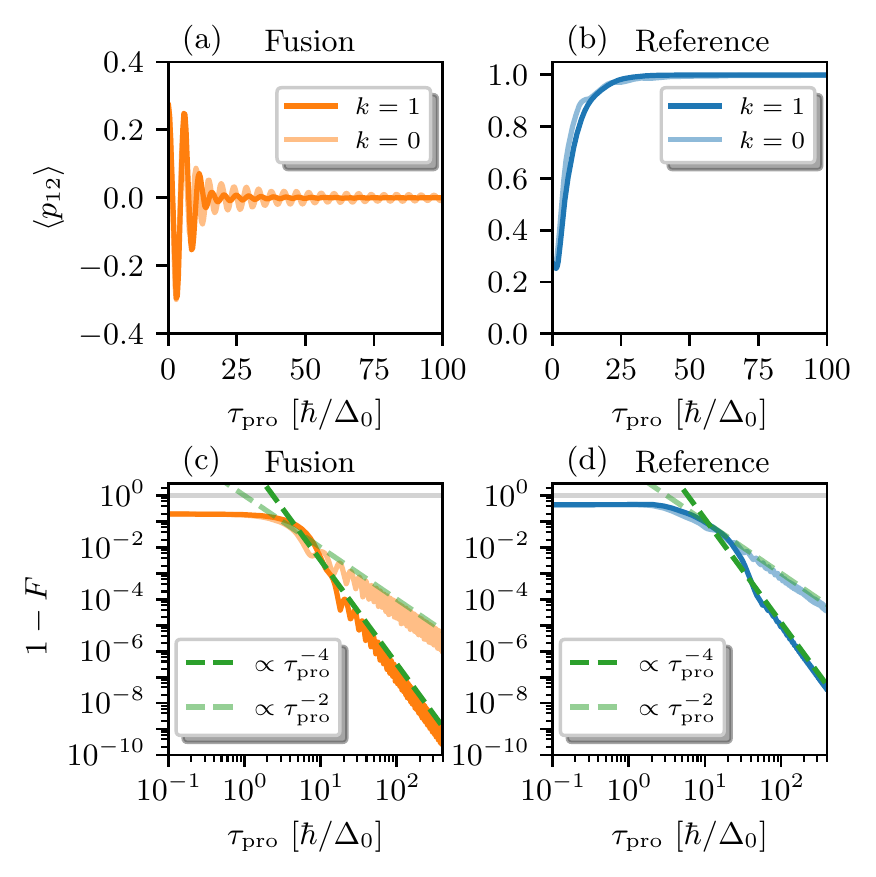}}
\caption{Fusion and reference protocols with diabatic errors. 
(a) and (b) Parity expectation $\langle \hat{p}_{12} \rangle $ as a function of protocol time $\tau_{\text{pro}}$.
(c) and (d) Infidelity $1-F$ as a function of $\tau_{\text{pro}}$.
Here, light-color curves use $f(x) = 1 - x$ with smoothness $k=0$, while normal-color ones use $f(x) = 1 - \sin^2(\pi x /2)$ with $k=1$.
}
\label{fig:diabatic}
\end{figure}

\section{Diabatic errors}
In realistic experiments, errors will inevitably occur due to imperfect quantum control or noise from the environment, both of which may blur the distinction between the outcomes of the fusion protocol relative to the reference one.
We first consider the diabatic errors because the fusion protocols have to be completed within a time scale shorter than the decoherence time of the Majorana system.
Our protocol is composed of two basic operations: 1. Tuning the dot energies from finite to zero, 2. Switching off the couplings between D2 and D3.
We assume that each operation takes half of the total protocol time, i.e., $\tau_{\text{pro}}/2$, and that the operations obey the same control function $f(x)$ which decreases monotonically from $f(0)=1$ to $f(1)=0$.
Here, $x=2\tau/\tau_{\text{pro}} \in [0,1]$ is the dimensionless time. 
As two concrete examples, we consider $f(x)=1-x$ with $k=0$ and $f(x)=1-\sin^2(\pi x/ 2)$ with $k=1$, where $k$ denotes the smoothness of Hamiltonian evolution, i.e., $H(\tau)$ is continuous and differentiable up to the first $k$ derivatives.
The dynamics of the system is governed by the time-dependent Schr\"odinger equation, and is calculated using the covariance matrix method~\cite{Kraus2010Generalized, Bravyi2017Complexity, Bauer2018Dynamics, Mishmash2020Dephasing}.
We will compute the expectation value of the parity $\hat{p}_{12}$, which is the quantity to be measured to verify the success of the fusion protocol and is defined as
\begin{align}\label{eq:parity}
    \langle \hat{p}_{12} \rangle \equiv \langle \psi_f | (i\hat{\gamma}_1 \hat{\gamma}_2)  |  \psi_f  \rangle,
\end{align}
where $\hat{\gamma}_1=\hat{c}_1 + \hat{c}\dg_1$ and $\hat{\gamma}_2=-i(\hat{c}_2 - \hat{c}\dg_2)$ denote the Majorana modes localized at dots D1 and D2, respectively, and $| \psi_f \rangle $ is the final state at $\tau=\tau_{\text{pro}}$.
In addition, we will compute a state infidelity, defined as
\begin{align}
    1 - F \equiv 1 - |\langle \phi | \psi_f  \rangle |^2,
\end{align}
where $| \phi \rangle $ is the target state in the idealized consideration. 
The infidelity, though difficult to measure, 
is a more precise metric of errors incurred in the protocol such as decoherence, which may not affect the parity outcome $p_{12}$.

Figures~\ref{fig:diabatic}(a) and~\ref{fig:diabatic}(b) show the numerically calculated parity expectation as a function of the protocol time.
For sufficiently long protocol time, $\langle \hat{p}_{12} \rangle$ approaches 0 and +1 for fusion and reference protocols, respectively, consistent with our analysis in the adiabatic limit. 
What's more, the convergence is reached faster for a smoother control function [see Fig.~\ref{fig:diabatic}(a) in particular].
Specifically, Fig.~\ref{fig:diabatic} shows that a protocol time of $\tau_{\text{pro}}\sim 25 \hbar/\Delta_0$ shows a parity $\langle \hat{p}_{12} \rangle$ which is quite close to zero in the fusion protocol compared to the reference value of one. 
This implies that a Majorana decay rate of approximately $5\%$ of the topological gap $\Delta_0$ (set by co-tunneling through the superconductor), if achieved in experiments, should allow for a convincing distinction between fusion and reference protocols.
Additionally, the infidelities for both protocols, as shown in Figs.~\ref{fig:diabatic}(c) and~\ref{fig:diabatic}(d), decay with the protocol time in a power-law fashion, i.e., $1- F \propto \tau^{-2k-2}_{\text{pro}}$, in the long-time limit, with its exponent depending only on the smoothness $k$, and not on other details.
Interestingly, the universal scaling behaviors of infidelity in anyon fusion are identical to those in anyon braiding or holonomy~\cite{Knapp2016Nature}.

\begin{figure*}[tbp]
\centering
\includegraphics[width= 6.8in]{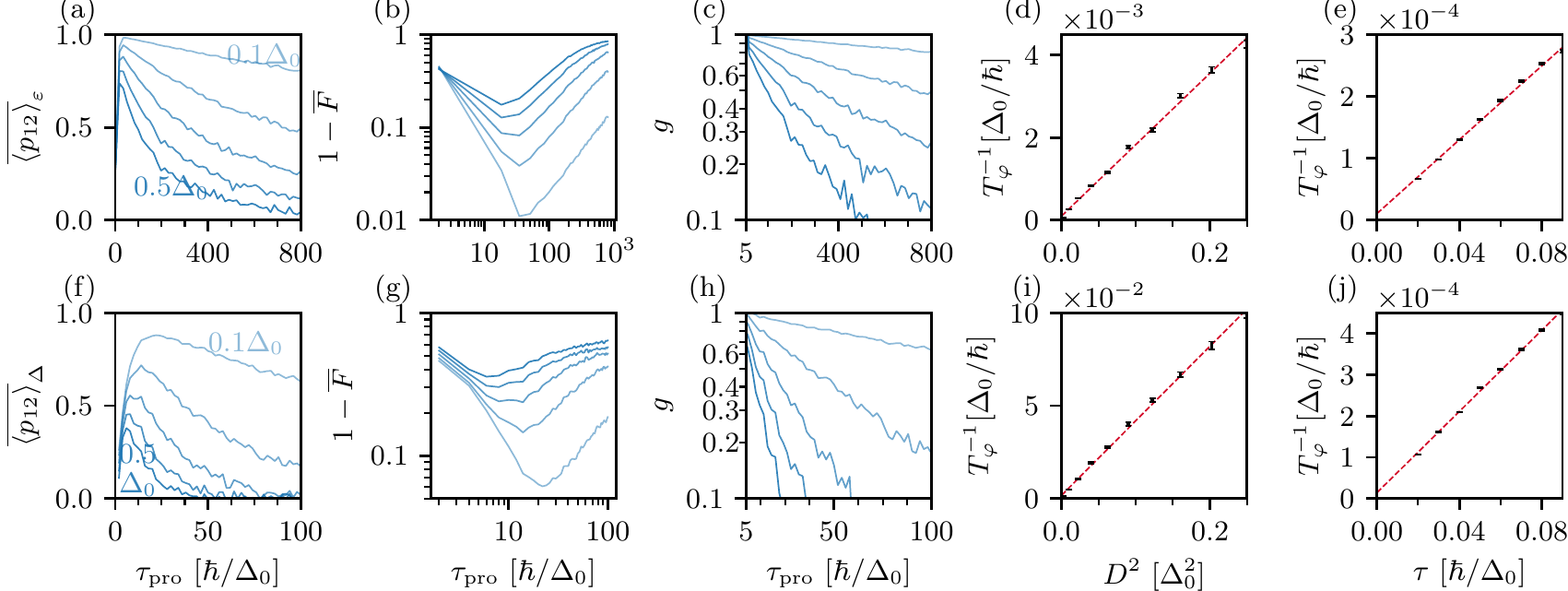}
\caption{
Dephasing effects due to noises in $\varepsilon_j$ (upper panels) and in $\Delta_{j,j+1}$ (lower panels).
(a) and (f): Noise-averaged parity versus $\tau_{\text{pro}}$ for $D=0.1\Delta_0$ (darkest blue) to $D=0.5\Delta_0$ (lightest blue) in the step of 0.1$\Delta_0$ and fixed correlation time $\tau_c=\hbar/\Delta_0$.
(b) and (g): Noise-averaged infidelity versus $\tau_{\text{pro}}$.
(c) and (h): Envelope function versus $\tau_{\text{pro}}$;
(d) and (i): Dephasing rate $T^{-1}_{\varphi}$ versus noise amplitudes $D^2$ at $\tau_c=\hbar/\Delta_0$. 
The black markers are extracted from (c) and (h) using Eq.~\eqref{eq:envelope}, and the dashed lines are linear fitting.
(e) and (j): Dephasing rate $T^{-1}_{\varphi}$ versus correlation time $\tau_c$ at $D^2=0.25\Delta^2_0$.
}
\label{fig:dephasing}
\end{figure*}

\section{Dephasing errors}
The Majorana decoherence process in the quantum dot chain, which was expected to limit the protocol time, is partly a result of fluctuation in the system parameters.  
In a realistic setup, these fluctuations are possibly induced by noises in the electrostatic gate voltages that control dot energies and effective couplings~\cite{Dvir2023Realization}.
Such noise combined with relaxation can lead to fluctuations of the Fermion occupation out of the ground state and has been shown to limit the fidelity of braiding, even in the case of ideal Majorana nanowires~\cite{Pedrocchi2015Majorana,Nag2019Diabatic}. 
The quantum dot chain is potentially more susceptible to such noise given its lack of robustness to parameter changes.
Here, we define $\lambda_{\alpha}(\tau)$ as the temporal fluctuation around the idealized value of a particular Hamiltonian parameter, and we assume that the fluctuations have zero mean and temporal correlations
\begin{align}
    \overline{\lambda_{\alpha}(\tau)} &= 0, \nn
    \overline{\lambda_{\alpha}(\tau) \lambda_{\beta}(\tau') } &= \delta_{\alpha \beta} S_{\alpha}(\tau-\tau'),
\end{align}
where $S_{\alpha}(\tau) = D^2_{\alpha} e^{-\tau^2 / (2\tau_{c, \alpha})^2}$ is the correlation function, with $D_{\alpha}$ the fluctuation amplitude and $\tau_{c, \alpha}$ the characteristic correlation time of $\lambda_{\alpha}(\tau)$.
We assume that the fluctuations in all the ten parameters of the Hamiltonian defined in Eq.~\eqref{eq:H_array} are completely independent of each other, and the final outcomes are averaged over 1000 different noise realizations.
Since the dephasing noise does not affect the fusion protocol in a significant way, we focus only on the reference protocol, leaving the discussion of the fusion protocol in the Supplemental material~\cite{supp_fusion}. 
The results of calibration errors from the ideal parameter values are also discussed in the Supplemental material~\cite{supp_fusion} and will be summarized at the end of the manuscript.

Figure~\ref{fig:dephasing}(a) shows the noise-averaged parity $\overline{\langle \hat{p}_{12} \rangle}$ as a function of protocol time with noise added only to the four dot energies.
In the long time limit, instead of approaching $+1$, the parity expectation now decays to zero due to random phase accumulation owing to noise, and the decay rate increases with the fluctuation amplitude $D_{\varepsilon}$.
The infidelity [see Fig.~\ref{fig:dephasing}(b)] is a non-monotonic function of protocol time, where the errors in the short and long protocol time regimes are dominated by diabatic and dephasing errors, respectively.
Interestingly, the effect of dephasing errors on the parity expectation is well described by an exponential decay envelope $g(\tau_{\text{pro}})$ [see Fig.~\ref{fig:dephasing}(c)] defined as
\begin{align}
g(\tau_{\text{pro}})  &\equiv \overline{\langle \hat{p}_{12} (\tau_{\text{pro}}) \rangle}_s/\langle \hat{p}_{12} (\tau_{\text{pro}}) \rangle \approx e^{ - \tau_{\text{pro}} / T_{\varphi} }
\label{eq:envelope}
\end{align}
where $s=\varepsilon, t$ or $\Delta$ denotes the type of parameter fluctuations, $\langle \hat{p}_{12} (\tau_{\text{pro}}) \rangle$ is subject to no noise [see Fig.~\ref{fig:diabatic}(b)], and $T_{\varphi}$ is the dephasing time.
In the weak fluctuation regime, the dephasing rate $T^{-1}_{\varphi}$ follows scaling behavior~\cite{supp_fusion}
\begin{align}\label{eq:dephasing}
    T^{-1}_{\varphi} \approx \sum_{\alpha} C_{\alpha} \tau_{c, \alpha} D^2_{\alpha},
\end{align}
where $C_{\alpha}$ is a proportionality constant.
Equation~\eqref{eq:dephasing} says that the dephasing rate is proportional to the correlation time $\tau_{c, \alpha}$ and variance $D^2_{\alpha}$, consistent with the numerical simulations shown in Figs.~\ref{fig:dephasing}(d) and~\ref{fig:dephasing}(e). 
These results validate the assumption in our analysis for diabatic errors, where we assume that the protocol time would be limited by dephasing.

To compare the effect of noises in dot energies with that in couplings, we repeat the same calculations, only including fluctuations in $\Delta_{j,j+1}$.
As shown in the lower panels of Fig.~\ref{fig:dephasing}, all the qualitative features discussed previously remain the same, but with a faster dephasing rate and hence larger infidelity.
This indicates that the Kitaev chain as well as the fusion protocols are more resilient against noises in dot energies than in coupling strengths.

\section{Parity readout}
\begin{figure}[t]
\centering
{\includegraphics[width = \linewidth]{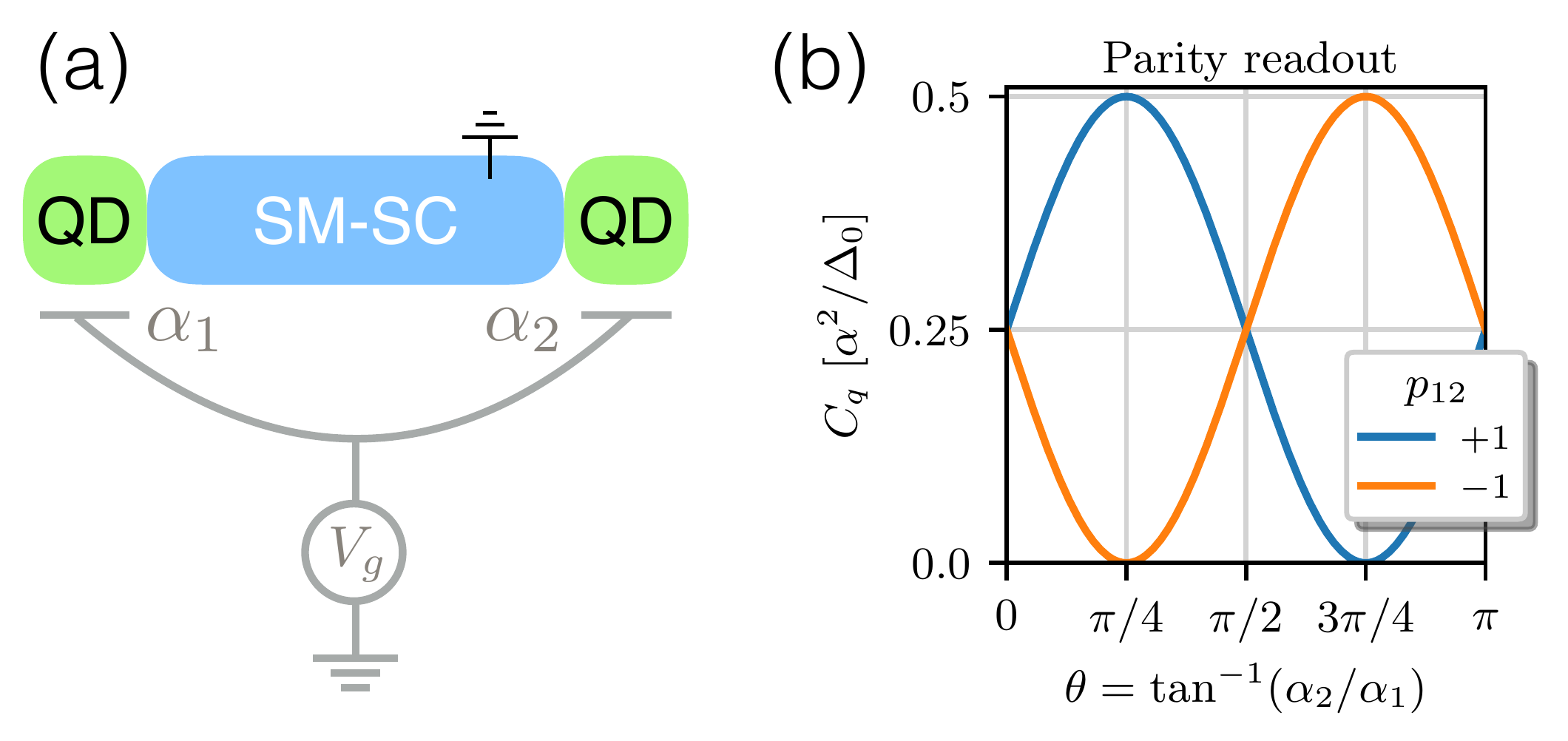}}
\caption{
(a) Schematic for parity readout on double quantum dots.
(b) Quantum capacitance $C_q$ for even- and odd-parity ground states at the sweet spot as a function of the relative lever arm strengths.
}
\label{fig:parity_readout}
\end{figure}

We finally discuss a readout scheme for the Fermion parity encoded in a pair of Majoranas, which would be used to determine the fusion outcomes.
Our scheme is based on measuring the quantum capacitance~\cite{Ruskov2019Quantum,Ruskov2021Modulated} of double quantum dots [see Fig.~\ref{fig:parity_readout}(a)], which measures the response of the even- and odd-parity ground states to gate voltage variations~\cite{Sillanpaa2005Direct, Esterli2019Small, Vigneau2022Probing}.
Here, we consider the Hamiltonian $\hat{H}_{\mathrm{DQD}} = \varepsilon_1 \hat{n}_1 + \varepsilon_2 \hat{n}_2 + t \hat{c}\dg_2 \hat{c}_1 + \Delta \hat{c}_2 \hat{c}_1 + \hc$.
Crucially, the two dot energies are controlled by a common electrostatic gate with generally different strengths of lever arms, i.e., $\varepsilon_{1,2} = \alpha_{1,2} \cdot V_g$.
At the sweet spot, the zero-temperature quantum capacitance is
\begin{align}
C_q = - \frac{\partial^2  E_{gs}}{ \partial V^2_g}
= \frac{1 + p_{12} \sin (2\theta)}{4} \cdot \frac{\alpha^2}{\Delta_0},
\label{eq:C_q}
\end{align}
where $p_{12}=\pm 1$ denotes the joint Fermion parity of $\gamma_{1,2}$ located on dots D1 and D2, $\alpha= \sqrt{\alpha^2_1 + \alpha^2_2}$ is the characteristic level arm strength, and $ \theta = \tan^{-1} (\alpha_2 / \alpha_1)$ denotes the ratio of two lever arm strengths.
As shown in Fig.~\ref{fig:parity_readout}(b), $C_q$ is a sinusoidal function of $\theta$, and has a $\pi/2$-phase shift between $p_{12}=\pm 1$, providing a different readout results for ground states with opposite parity.
In particular, the readout visibility is maximal at $\theta=\pi/4$ or $3\pi/4$, where the two lever arms are equal in strength.
By contrast, at $\theta=0$ or $\pi/2$ corresponding to $\alpha_2=0$ or $\alpha_1=0$, respectively, the two parity states become indistinguishable at the sweet spot. 
With one of the lever arms strength being zero, the quantum capacitance measurement is on only one dot and therefore is incapable of reading out the \emph{nonlocal} parity information encoded in two dots. As shown in the Supplementary material~\cite{supp_fusion}, the measurement results in Fig.~\ref{fig:parity_readout} are robust against substantial calibration errors in parameters.

\section{Discussion and summary}
In this work, we give concrete protocols for detecting the Majorana fusion rules in quantum dots.
Manipulation and readout of Majoranas are implemented in a fully electrostatic way,  i.e., either the dot energies and the effective coupling strengths can be tuned by varying the voltage of the electrostatic gate individually~\cite{Dvir2023Realization}.
Removing the need of superconducting islands makes our proposal particularly relevant and suitable for the ongoing experimental efforts~\cite{Dvir2023Realization}.
Using numerical simulations, we show that diabatic and dephasing errors altogether set the constraint for the protocol time, i.e., the operations should neither be too fast to break the adiabatic condition nor too slow to accumulate dephasing errors.
In this aspect, although not directly demonstrating the fusion rules, the reference protocol is of paramount importance in extracting the dephasing time and in excluding the potentially false-positive interpretations of the fusion results~\cite{Clarke2017Probability}.
Specifically, we find that demonstration of the fusion protocol should be feasible even for a $10\%$  noise-induced variation (or calibration errors) in parameters.
We also propose a quantum capacitance measurement of the parity encoded in double quantum dots, which is applicable to either dot D1 and D2, or dot D3 and D4, eliminating the need of an extra quantum dot in the charge-parity transfer method~\cite{Flensberg2011Nonabelian, Krojer2022Demonstrating}.
Taking the parameter values from Ref.~\cite{Dvir2023Realization}, we estimate that $C_q \sim \alpha^2/\Delta_0 \sim 0.75$ fF, with $\alpha \approx 0.3~e$, and $\Delta_0 \approx 20~\mu$eV, within the reach of the state-of-the-art measurement techniques~\cite{Malinowski2022Radio, deJong2021Rapid, Ibberson2021Large, Petersson2010Charge}.

Since this architecture is the minimal setup for realizing a Majorana qubit in an artifical Kitaev chain, the first measurement would be the Majorana coherence time. This can be measured using the parity readout scheme we propose by looking for photon assisted tunneling or Rabi oscillations~\cite{vanZanten2020Photon} in the parity of the chain in the four Majorana configuration (i.e. F2 or R2) where the middle link is partially cut. The fusion measurement we propose could then be done as long as the decoherence time $T_\varphi$ is an order of magnitude longer than the Rabi oscillation period which is on the scale $\hbar/\Delta_0\sim0.03$~ns (i.e., the time for convincing demonstration of Rabi oscillations in a Majorana qubit). 

\begin{acknowledgements}
We are grateful to Tom Dvir, Guanzhong Wang, Filip K. Malinowski, Srijit Goswami, Christian Prosko and Anton Akhmerov for useful discussions. F. Setiawan made his technical contributions while at the University of Chicago.
This work was supported by a subsidy for Top Consortia for Knowledge and Innovation (TKl toeslag), 
by the Laboratory for Physical Sciences through the Condensed Matter Theory Center, by University of Maryland High-Performance Computing Cluster (HPCC), by National Science Foundation (Platform for the Accelerated Realization, Analysis, and Discovery of Interface Materials (PARADIM)) under Cooperative Agreement No. DMR-1539918, by the Army Research Office under Grant Number W911NF-19-1-0328.

\end{acknowledgements}

\bibliography{references.bib}

\onecolumngrid
\vspace{1cm}
\begin{center}
{\bf\large Supplemental Materials for ``Fusion protocol for Majorana modes in coupled quantum dots"}
\end{center}
\vspace{0.5cm}

\setcounter{secnumdepth}{3}
\setcounter{equation}{0}
\setcounter{figure}{0}
\renewcommand{\theequation}{S-\arabic{equation}}
\renewcommand{\thefigure}{S\arabic{figure}}
\renewcommand\figurename{Supplementary Figure}
\renewcommand\tablename{Supplementary Table}
\newcommand\Scite[1]{[S\citealp{#1}]}
\newcommand\Scit[1]{S\citealp{#1}}

\makeatletter \renewcommand\@biblabel[1]{[S#1]} \makeatother

\section{Numerical approach to time-dependent Schr\"odinger equation}\label{App:Time-dep}

In this section, we briefly introduce the details of our numerical method to solve the time-dependent Schr\"odinger equation. We construct the covariance matrix formalism to describe the system in terms of the Majorana operations~\cite{Bravyi2017Complexity}, i.e., 
\begin{equation}\label{eq:Gamma} 
    \hat{\Gamma}_{ij}=\frac{i}{2}\expval{\left[ \hat{\chi}_{i} , \hat{\chi}_{j}  \right]},
\end{equation}
where $[,]$ means the commutation of two operators, and $\expval{\dots}$ takes the expected value with respect to the ground state.
$\chi_i$ are Majorana operators defined as $\chi_{2n-1} = c_n + c\dg_n$ and $\chi_{2n} = i(c_n - c\dg_n)$. 

In our case, the ground states of two types of protocols are Eq.~(3) for the fusion protocol, and Eq.~(4) for the reference protocol, which gives covariance matrices following Eq.~\eqref{eq:Gamma} as per
\begin{equation}
    \hat{\Gamma}_{F}=\begin{pmatrix}
        0& 0& 0& 0& 0& 0& 0& -1\\
        0& 0& 0& 0& 1& 0& 0& 0\\
        0& 0& 0& 0& 0& 1& 0& 0\\
        0& 0& 0& 0& 0& 0& 1& 0\\
        0& -1& 0& 0& 0& 0& 0& 0\\
        0& 0& -1& 0& 0& 0& 0& 0\\
        0& 0& 0& -1& 0& 0& 0& 0\\
        1& 0& 0& 0& 0& 0& 0& 0\\
    \end{pmatrix} , 
    \hat{\Gamma}_{R}=\begin{pmatrix}
        0& 0& 0& 0& 0& -1& 0& 0\\
        0& 0& 0& 0& 1& 0& 0& 0\\
        0& 0& 0& 0& 0& 0& 0& -1\\
        0& 0& 0& 0& 0& 0& 1& 0\\
        0& -1& 0& 0& 0& 0& 0& 0\\
        1& 0& 0& 0& 0& 0& 0& 0\\
        0& 0& 0& -1& 0& 0& 0& 0\\
        0& 0& 1& 0& 0& 0& 0& 0 \\
    \end{pmatrix}.
\end{equation}

Therefore, the time evolution of the covariance matrix $\hat{\Gamma}$ under the Hamiltonian $\hat{H}$ follows the time-dependent Schr\"odinger equation 
\begin{equation}\label{eq:Sch}
    \hbar\frac{d \hat{\Gamma}}{d\tau}=\left[\hat{A},\hat{\Gamma}\right],
\end{equation}
where $\hat{A}$ corresponds to the matrix element of the Hamiltonian in Majorana basis, i.e.,
\begin{equation}
    \hat{H}_M=\frac{i}{4} \sum_{i,j}A_{ij} \hat{\chi}_{i} \hat{\chi}_{j}.
\end{equation}

To solve the differential equation Eq.~\eqref{eq:Sch}, we flatten the covariance matrix into a vector, and feed it into a standard ODE solver. To allow the adaptive sampling of the time in solving Eq.~\eqref{eq:Sch}, we interpolate the disorder profile in the time domain.

\section{Fluctuation in the normal coupling $t$ between the quantum dots}\label{App:t}
\begin{figure}
    \centering
    \includegraphics[width=6.8in]{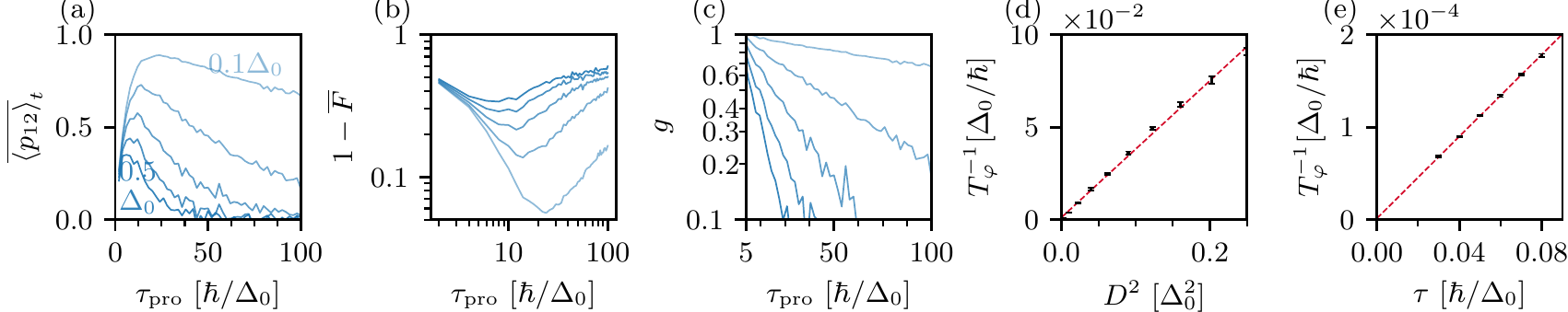}
    \caption{Dephasing errors for fluctuations in normal interdot coupling strength $t$. 
    (a) Ensemble average parity versus $\tau_{\text{pro}}$ for $D=0.1\Delta_0$ (darkest blue) to $D=0.5\Delta_0$ (lightest blue) in the step of 0.1$\Delta$ and $\tau=\hbar/\Delta$;
    (b) Ensemble average error versus $\tau_{\text{pro}}$; 
    (c) Ensemble average envelope versus $\tau_{\text{pro}}$;
    (d) $T^{-1}_2$ versus strength $D^2$ at $\tau=\hbar/\Delta$. The black markers show the numerical results in an ensemble of 1000 along with the error bars showing the standard error. The blue dashed lines show the fitted slopes and interceptions, which is $(0.37~\hbar/\Delta^3, 1.27\times 10^{-3}\Delta_0/\hbar)$;
    (e) $T^{-1}_2$ versus correlation time $\tau$ at $D=0.5\Delta_0$. The fitted slopes and interceptions are $(1.8\times10^{-2}\Delta_0^2/\hbar^2, 6.2\times 10^{-5}\Delta_0/\hbar)$. 
    }
    \label{fig:dephasing_t}
\end{figure}

In this section, we study the dephasing noise of the normal coupling $t$ in a similar way as we did in Fig.~3. From Figs.~\ref{fig:dephasing_t}(a-c), we also present the parity, wave function error, and decaying envelope, which all show a similar exponential decay as the results from the fluctuation in $\Delta$ as shown in Fig.~3. In particular, in Figs.~\ref{fig:dephasing_t}(d-e), we fit the decay exponent of the envelope and find similar slopes and intercepts as in Figs.~3(i-h), which indicates that the response of this protocol to the fluctuation of $t$ and $\Delta$ are qualitatively the same, and both of them have much significant adverse impact than the fluctuation in $\varepsilon$.

\section{Analytic derivation of dephasing time $T_{\varphi}$}

In this section, we analytically derived the dephasing time $T_\varphi$ in Eq.~(8). Our model resembles the concept of linear tetron in Ref.~\onlinecite{karzig2017scalable}, where the computation basis are defined as
\begin{equation}
    \begin{split}
        \ket{0}&=\ket{0_{12},0_{34}}=\ket{p_{12}=p_{34}=1}\\
        \ket{1}&=\ket{1_{12},1_{34}}=\ket{p_{12}=p_{34}=-1}
    \end{split}
\end{equation}
because we adopt the convention that fixes the total parity to be even ($p_{12}p_{34}=1$). 
Therefore, the Pauli matrix on each qubit in the terms of the Majorana operators are
\begin{equation}
    \begin{split}
        X&=i \hat{\gamma}_{2} \hat{\gamma}_{3} =i \hat{\gamma}_{1} \hat{\gamma}_{4} ,\\
        Y&=i \hat{\gamma}_{1} \hat{\gamma}_{3} =-i\hat{\gamma}_{2} \hat{\gamma}_{4} ,\\
        Z&=i \hat{\gamma}_{1} \hat{\gamma}_{2} = i \hat{\gamma}_{3}\hat{\gamma}_{4} 
    \end{split}
\end{equation} 
up to an overall phase.
In the long time limit (i.e., near the end of the protocol time), we can approximate the Hamiltonian in Eq.~(2) as 
\begin{equation}
    \begin{split}
        \hat{H}(\tau)&=i\varepsilon_{12}(\tau) \hat{\gamma}_{1} \hat{\gamma}_{2} + i \varepsilon_{23}(\tau) \hat{\gamma}_{2} \hat{\gamma}_{3} + i \varepsilon_{34}(\tau) \hat{\gamma}_{3} \hat{\gamma}_{4} \\
        &= \left[ \varepsilon_{12}(\tau)+\varepsilon_{34}(\tau) \right]\hat{Z} + \varepsilon_{23}(\tau)\hat{X},
    \end{split}
\end{equation}
where $\varepsilon_{i,i+1}(\tau)$ can be obtained by calculating the overlap between two Majorana operators at quantum dot $i$ and $i+1$, and the wavefunction of the reference state as $\ket{\psi_i}_R=\ket{0}$. Note that the subscript $i$ here indicates the initial state, which looks contradictory to Eq.~(4). However, this is intended as we only study the behavior of the long-time limit, where the MZMs have already formed. Equivalently, this means we can shift the starting time from $\tau=0$, where the system is in the atomic limit, to $\tau=\tau_0\lesssim \tau_{\mathrm{pro}}$. This approximation will not affect the long-time behavior of the decaying envelope.

Therefore, the time evolution of the parity operator $\hat{p}_{12}(\tau)$ for the reference state is 
\begin{equation}
    \begin{split}
        \hat{p}_{12}(\tau)&=e^{i\int_{\tau_0}^{\tau}\hat{H}(\tau^\prime) d\tau^\prime} \hat{p}_{12}(\tau_0) e^{-i\int_{\tau_0}^{\tau}\hat{H}(\tau^\prime) d\tau^\prime}\\
        &=e^{i \left[ f_Z(\tau) \hat{Z} + f_X(\tau) \hat{X} \right] } \hat{X} e^{-i \left[ f_Z(\tau) \hat{Z} + f_X(\tau) \hat{X} \right] }\\
        &=\frac{f_{X}(\tau) f_{Z}(\tau) \left( 1-\cos(2\sqrt{f_{X}^2(\tau)+ f_{Z}^2(\tau)}) \right)}{f_{X}^2(\tau)+f_{Z}^2(\tau)}\hat{X}+ \frac{f_{X}(\tau) \sin(2\sqrt{f_{X}^2(\tau)+f_{Z}^2(\tau)})}{f_{X}^2(\tau)+f_{Z}^2(\tau)} \hat{Y} \\
        &+\frac{f_{Z}^2(\tau)+f_{X}^2(\tau)\cos(2\sqrt{f_{X}^2(\tau)+ f_{Z}^2(\tau)})}{f_{X}^2(\tau)+f_{Z}^2(\tau)}\hat{Z},
    \end{split}
\end{equation} 
where $f_X(\tau)=\int_{\tau_0}^\tau \varepsilon_{23}(\tau^\prime) d\tau^\prime$, and $f_Z(\tau)=\int_{\tau_0}^\tau \varepsilon_{12}(\tau^\prime) + \varepsilon_{34}(\tau^\prime) d\tau^\prime$. Thus, the final parity of the reference state is 
\begin{equation}
    \begin{split}
        \bra{\psi_f}{\hat{p}_{12}}\ket{\psi_f}_R&=\bra{\psi_i}{\hat{p}_{12}(\tau)}\ket{\psi_i}_R=\frac{f_{Z}^2(\tau)+f_{X}^2(\tau)\cos(2\sqrt{f_{X}^2(\tau)+ f_{Z}^2(\tau)})}{f_{X}^2(\tau)+f_{Z}^2(\tau)}\\
        &=1-\frac{2f_{X}^2(\tau) \sin^2{\sqrt{f_{X}^2(\tau)+f_{Z}^2(\tau)  }}}{f_{X}^2(\tau)+f_{Z}^2(\tau) },
    \end{split}
\end{equation}
and the disorder-averaged parity is 
\begin{equation}\label{eq:sum}
    \begin{split}
        \overline{\bra{\psi_f}{\hat{p}_{12}}\ket{\psi_f}_R}&=1-\overline{\frac{2f_{X}^2(\tau) \sin^2{\sqrt{f_{X}^2(\tau)+f_{Z}^2(\tau)  }}}{f_{X}^2(\tau)+f_{Z}^2(\tau) }}\\
        &\approx \sum_{n=0} \overline{\left( \frac{i \sqrt{2}f_{X}(\tau) \sin{\sqrt{f_{X}^2(\tau)+f_{Z}^2(\tau)  }}}{\sqrt{f_{X}^2(\tau)+f_{Z}^2(\tau)}} \right)^n}\\
        &=\overline{\exp(\frac{i \sqrt{2}f_{X}(\tau) \sin{\sqrt{f_{X}^2(\tau)+f_{Z}^2(\tau)  }}}{\sqrt{f_{X}^2(\tau)+f_{Z}^2(\tau)}})}
    \end{split}
\end{equation}
where the second line holds because we can expand $\varepsilon_{23}(\tau)$ to the second order of disorder $\delta\lambda_\alpha(\tau)$ as
\begin{equation}\label{eq:ep23}
    \varepsilon_{23}(\tau) \approx \varepsilon_{23}^{(0)}(\tau) + \sum_{\alpha} \delta\lambda_\alpha(\tau) \dot{\varepsilon}_{23,\alpha}(\tau) + \frac{1}{2} \sum_{\alpha,\beta} \delta\lambda_\alpha(\tau) \delta\lambda_\beta(\tau) \ddot{\varepsilon}_{23,\alpha,\beta}(\tau),
\end{equation} 
where 
\begin{equation}\label{eq:derivative}
    \dot{\varepsilon}_{23,\alpha}(\tau)=\eval{ \frac{\partial \varepsilon_{23,\alpha}(\tau)}{\partial \lambda_\alpha(\tau)} }_{\delta\lambda_\alpha=0},\quad
    \ddot{\varepsilon}_{23,\alpha,\beta}(\tau)=\eval{\frac{\partial^2 \varepsilon_{23,\alpha}(\tau)}{\partial \lambda_\alpha(\tau)^2}}_{\delta\lambda_\alpha=0},
\end{equation}
and $\varepsilon_{23}^{(0)}(\tau)$ is a deterministic function of time without disorder. Because we define the disorder to be zero mean, $\overline{\delta\lambda_\alpha(\tau)}=0$ , the term $\overline{\delta\lambda_\alpha(\tau) \dot{\varepsilon}_{23,\alpha}(\tau)}$ vanishes up to $O(\delta\lambda(\tau))$, and the higher order terms in Eq.~\eqref{eq:ep23} can be absorbed into the later part in the summation with $n\ge2$ in Eq.~\eqref{eq:sum}. The remaining term is the first term in Eq.~\eqref{eq:ep23} with a constant $\int_{\tau_0}^\tau \varepsilon_{23}^{(0)}(\tau^\prime) d\tau^\prime$, which, however, could be treated as zero because the coupling between the second quantum dot and the third quantum has already been turned off near the final stage of the protocol for the reference protocol, which indicates $f_X(\tau) \ll f_Z(\tau)$. 

Therefore, the exponent in Eq.~\eqref{eq:sum} can be further simplified to
\begin{equation}\label{eq:sinfz}
    \begin{split}
        & i \sqrt{2} f_{X}(\tau) \frac{\sin(f_{Z}(\tau)) }{f_{Z}(\tau) }\\
        =&i \sqrt{2} f_{X}(\tau) \frac{\sin(\int_{\tau_0}^{\tau}  \left[ \varepsilon_Z^{(0)}(\tau^\prime)+\sum_{\alpha} \dot{\varepsilon}_{Z,\alpha}(\tau^\prime) \lambda_\alpha(\tau^\prime) \right] d\tau^\prime ) }{\int_{\tau_0}^\tau  \left[ \varepsilon_Z^{(0)}(\tau^\prime)+\sum_{\alpha} \dot{\varepsilon}_{Z,\alpha}(\tau^\prime) \lambda_\alpha(\tau^\prime) \right] d\tau^\prime}\\
        \approx& i \sqrt{2} f_{X}(\tau) \frac{\sin(\int_{\tau_0}^\tau \varepsilon_Z^{(0)}(\tau^\prime) d\tau^\prime)\cos(\sum_{\alpha}\int_{\tau_0}^\tau \dot{\varepsilon}_{Z,\alpha}(\tau^\prime)\lambda_\alpha(\tau^\prime)  d\tau^\prime)+\cos(\int_{\tau_0}^\tau \varepsilon_Z^{(0)}(\tau^\prime) d\tau^\prime) \sin(\sum_{\alpha}\int_{\tau_0}^\tau \dot{\varepsilon}_{Z,\alpha}(\tau^\prime)\lambda_\alpha(\tau^\prime)  d\tau^\prime) }{\int_{\tau_0}^\tau  \left[ \varepsilon_Z^{(0)}(\tau^\prime)\right] d\tau^\prime}
    \end{split}
    \end{equation}
where $\varepsilon_Z(\tau)=\varepsilon_{12}(\tau)+\varepsilon_{34}(\tau)$, and $\dot{\varepsilon}_{Z,\alpha}(\tau)$ is defined in the similar way in Eq.~\eqref{eq:derivative}. Since $\int_{\tau_0}^\tau \dot{\varepsilon}_{Z,\alpha}(\tau^\prime)\delta\lambda_\alpha(\tau^\prime)  d\tau^\prime$ is proportional to the small disorder $\delta\lambda_\alpha(\tau)$, we take the approximation of $\cos(\sum_{\alpha}\int_{\tau_0}^\tau \dot{\varepsilon}_{Z,\alpha}(\tau^\prime)\delta\lambda_\alpha(\tau^\prime)  d\tau^\prime)\approx 1$, and $\sin(\sum_{\alpha}\int_{\tau_0}^\tau \dot{\varepsilon}_{Z,\alpha}(\tau^\prime)\delta\lambda_\alpha(\tau^\prime)  d\tau^\prime)\approx \sum_{\alpha}\int_{\tau_0}^\tau \dot{\varepsilon}_{Z,\alpha}(\tau^\prime)\delta\lambda_\alpha(\tau^\prime)  d\tau^\prime$. Furthermore, because $f_X $ already contains the leading order of $\delta\lambda_\alpha(\tau^\prime)$ as shown in Eq.~\eqref{eq:ep23}, the second term on the numerator of Eq.~\eqref{eq:sinfz} can be omitted up to $O(\delta\lambda_\alpha(\tau^\prime))$. Therefore, the disorder averaged parity Eq.~\eqref{eq:sum} becomes simple
\begin{equation}\label{eq:exp}
        \overline{\bra{\psi_f}{\hat{p}_{12}}\ket{\psi_f}_R}=\overline{\exp(i k f_{X}(\tau) )}=\overline{\exp(i k \int_{\tau_0}^\tau \varepsilon_{23}(\tau^\prime) d\tau^\prime )},        
\end{equation}  
where the constant $k=\sqrt{2}\dfrac{\sin({\int_{\tau_0}^\tau f_{X}(\tau^\prime) \varepsilon_Z^{(0)}(\tau^\prime) d\tau^\prime})}{{\int_{\tau_0}^\tau f_{X}(\tau^\prime) \varepsilon_Z^{(0)}(\tau^\prime) d\tau^\prime}}$. 

The next step is to evaluate the Eq.~\eqref{eq:exp} and extract the dephasing time $T_\varphi$. We follow the similar steps in Ref.~\onlinecite{Mishmash2020Dephasing}, and first discretize the time to obtain
\begin{equation}
    \begin{split}
        \overline{\exp(i k \int_{\tau_0}^\tau \varepsilon_{23}(\tau^\prime) d\tau^\prime  )}&= \frac{1}{\mathcal{Z}}\int \mathcal{D}\delta\lambda_\alpha(\tau) e^{i k  \Delta\tau \sum_{\tau^\prime=\tau_0}^{\tau} \varepsilon_{23}(\tau^\prime)} e^{- \frac{1}{2} \sum_{\tau_1,\tau_2}\sum_{\alpha} S^{-1}_{\alpha}(\tau_1-\tau_2)\delta\lambda_{\alpha}(\tau_1)\delta\lambda_{\alpha}(\tau_2)}\\
        &=\frac{ e^{i k \int_{\tau_0}^\tau \varepsilon_{23}^{(0)}(\tau^\prime) d\tau^\prime} }{\mathcal{Z}} \int \mathcal{D}\delta\lambda_\alpha(\tau) e^{i k  \Delta \tau \sum_{\tau^\prime=\tau_0}^{\tau}  \sum_{\alpha} \delta\lambda_\alpha(\tau^\prime) \dot{\varepsilon}_{23,\alpha}(\tau^\prime) } e^{- \frac{1}{2} \sum_{\tau_1,\tau_2}\sum_{\alpha} S^{-1}_{\alpha}(\tau_1-\tau_2)\delta\lambda_{\alpha}(\tau_1)\delta\lambda_{\alpha}(\tau_2)}
    \end{split}
\end{equation}
where 
\begin{equation}
    \mathcal{Z}= \int \mathcal{D}\delta\lambda_\alpha(\tau) e^{-\frac{1}{2} \sum_{\tau_1,\tau_2}\sum_{\alpha} S^{-1}_{\alpha}(\tau_1-\tau_2)\delta\lambda_{\alpha}(\tau_1)\delta\lambda_{\alpha}(\tau_2)}.
\end{equation}

From the n-dimensional Gaussian integral, we integrate out $\mathcal{D}\lambda_\alpha(t)$ as
\begin{equation}
    \overline{\exp(i k \int_{\tau_0}^\tau \varepsilon_{23}(\tau^\prime) d\tau^\prime  )}= e^{i k \int_{\tau_0}^\tau \varepsilon_{23}^{(0)}(\tau^\prime) d\tau^\prime} e^{\frac{1}{2} \sum_{\alpha} \sum_{\tau_1,\tau_2}\left[ ik\Delta\tau\dot{\varepsilon}_{23,\alpha}(\tau_1) \right] S_\alpha(\tau_1-\tau_2)\left[ ik\Delta\tau\dot{\varepsilon}_{23,\alpha}(\tau_2) \right]},
\end{equation} 
where the exponent of the last term above can be integrated in the continuous limit as
\begin{equation}
    \begin{split}
        &\frac{1}{2} \sum_{\alpha} \sum_{\tau_1,\tau_2}\left[ ik\Delta\tau\dot{\varepsilon}_{23,\alpha}(\tau_1) \right] S_\alpha(\tau_1-\tau_2)\left[ ik\Delta\tau\dot{\varepsilon}_{23,\alpha}(\tau_2) \right]\\
        =&-\sum_{\alpha} k^2\int_{\tau_0}^{\tau}\int_{\tau_0}^{\tau} d\tau_1 d\tau_2 S_\alpha(\tau_1-\tau_2) \dot{\varepsilon}_{23,\alpha}(\tau_1)\dot{\varepsilon}_{23,\alpha}(\tau_2)\\
        \approx& - \sum_\alpha(\dot{\varepsilon}_{23,\alpha})^2 k^2 \int_{\tau_0}^{\tau}\int_{\tau_0}^{\tau} d\tau_1 d\tau_2 D^2_{\alpha} e^{-(\tau_1-\tau_2)^2 / (2\tau_{c, \alpha})^2}\\
        \approx& - \sum_\alpha(\dot{\varepsilon}_{23,\alpha})^2 k^2 D^2_{\alpha} 2 \sqrt{\pi}  \tau_{c,\alpha} \tau,
    \end{split}
\end{equation}
where $\dot{\varepsilon}_{23,\alpha}(\tau^\prime)$ is considered to be slowly varying such as it can be replaced by the averaged value $\dot{\varepsilon}_{23,\alpha}=\frac{1}{\tau-\tau_0}\int_{\tau_0}^\tau\dot{\varepsilon}_{23,\alpha}(\tau^\prime) d\tau^\prime$, and the last line is reached in the long time limit, i.e., $\tau\gg\tau_{c,\alpha}$. Therefore, we recover the dephasing rate in Eq.~(9) as
\begin{equation} 
    T_{\varphi}^{-1}=\sum_\alpha(\dot{\varepsilon}_{23,\alpha})^2 k^2 D^2_{\alpha} 2 \sqrt{\pi}  \tau_{c,\alpha}  = \sum_\alpha C_\alpha \tau_{c,\alpha} D_{\alpha}^2,
\end{equation}
with $C_\alpha$ being a constant which can be determined from the fitting.

\section{Calibration errors}

\subsection{Fusion protocols}
\begin{figure*}[ht]
    \centering
    \includegraphics[width=6.8in]{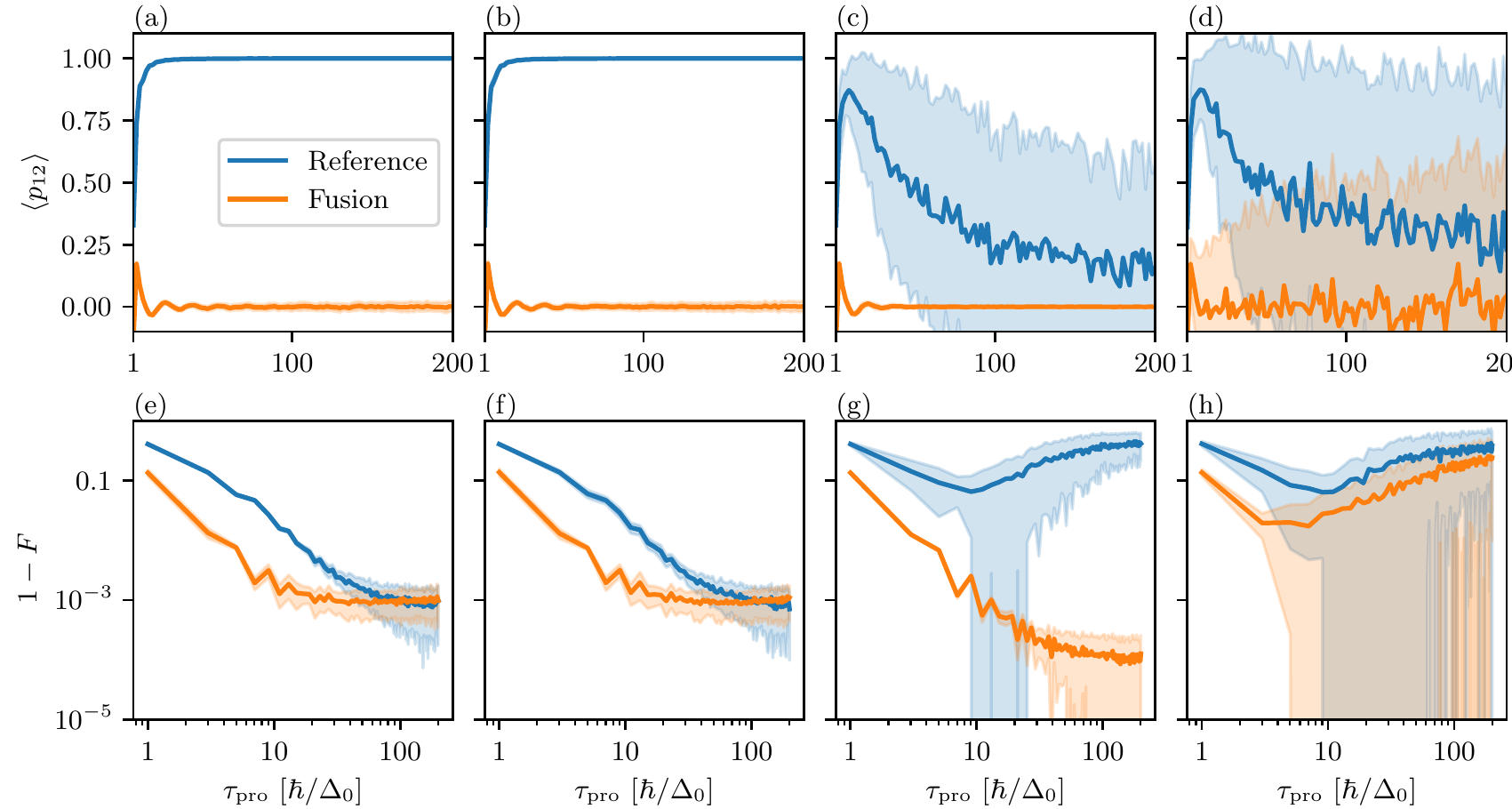}
    \caption{Parity (upper panel) and fidelity (lower panel) as a function of $\tau_{\mathrm{pro}}$ with the error amplitude of $0.1\Delta_0$ for (a,e) $t_{i,i+1}$, $\Delta_{i,i+1}$, and $E_i$ are perfectly calibrated initially while only $E$ has calibration error in the final state; (b,f) $t_{i,i+1}$ and $\Delta_{i,i+1}$ are perfectly calibrated initially while $E_i$ has calibration error both in the initial stage and final stage; (c,g) $t_{i,i+1}$, $\Delta_{i,i+1}$, and $E_i$ are perfectly calibrated throughout the protocol while the calibration error happens only to $t_{23}$ at the final state; (d,h) $E_i$ are perfectly calibrated throughout the protocol while the calibration error happens to all $t_{i,i+1}$ and $\Delta_{i,i+1}$.}
    \label{fig:calibration_error}
\end{figure*}

In this section, we study the calibration error while tuning the parameters of $\varepsilon$, $t$, and $\Delta$. The error is simulated using a uniform distribution within $\pm 0.1\Delta_0$ independently for each parameter. We consider four situations as shown in Fig.~\ref{fig:calibration_error}.

In Figs.~\ref{fig:calibration_error}(a, e), calibration errors only take place in the four dot energies $\varepsilon_{i}$ at the final state. The effects on parity expectation and infidelity are both minimal.
In Figs.~\ref{fig:calibration_error}(b,f), calibration errors only take place in the four dot energies $\varepsilon_{i}$, but at both the initial and the final states. 
Similar to Fig.~\ref{fig:calibration_error}(a,e), the effects are also minimal.
The results in the first two columns indicate that calibration errors in dot energies have minor effects on the fusion outcome.
In Figs.~\ref{fig:calibration_error}(c, g), calibration errors only take place in only $t_{23}$ and $\Delta_{23}$ in the final state of the switch-off. 
Interestingly, it has a much more adverse effect on the reference protocol than the fusion one.
This is because in the reference one, we first try to switch off $t_{23}$ and $\Delta_{23}$ but with calibration errors, introducing large parity-breaking errors in the second step of the protocol.
By contrast, calibration errors appear only towards the very end of the whole protocol in the fusion one, mitigating the effect of calibration errors.
In Figs.~\ref{fig:calibration_error}(d, h), calibration errors in $t_{i,i+1}$ and $\Delta_{i,i+1}$ appear in both initial and final states, giving an adverse effects on both fusion and reference protocols.
To summarize, calibration errors in couplings between dots have a more adverse effect on the fusion outcome then errors in the on-site energies.
Particularly, calibration errors in couplings in the initial states is more adverse than those in the final states.

\subsection{Parity readout}
\begin{figure}
    \centering
    \includegraphics[width=6.8in]{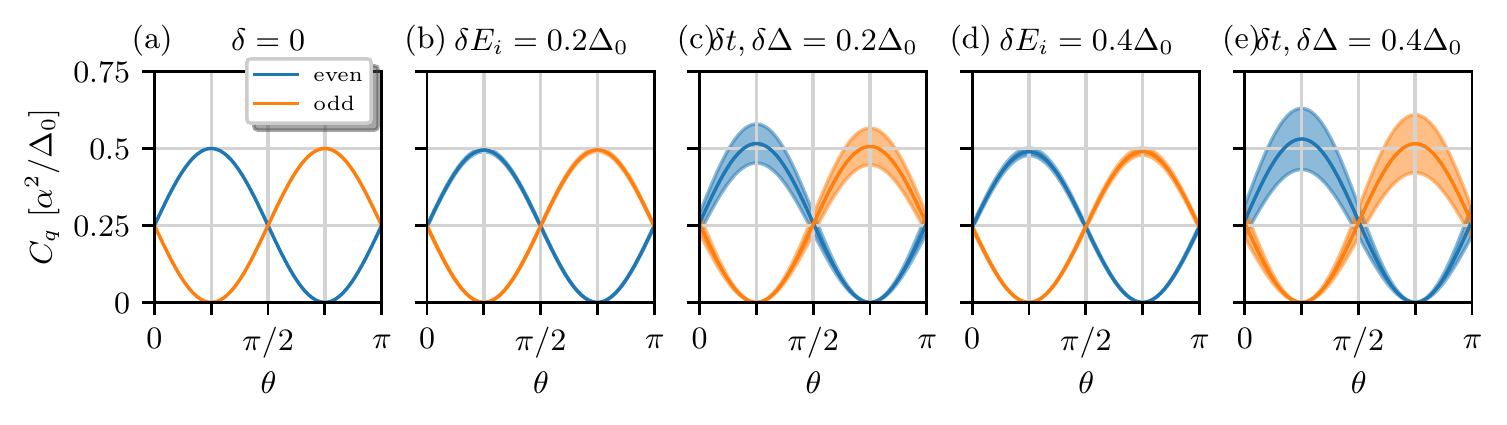}
    \caption{Parity readout in the presence of calibration errors.
    (a) $C_q$ without calibration errors.
    (b) Calibration errors are only in $E_L$ and $E_R$ with error amplitude $\delta=0.2~\Delta_0$.
    (c) Calibration errors are only in $t$ and $\Delta$ with error amplitude $\delta=0.2~\Delta_0$.
    (d) Calibration errors are only in $E_L$ and $E_R$ with error amplitude $\delta=0.4~\Delta_0$.
    (e) Calibration errors are only in $t$ and $\Delta$ with error amplitude $\delta=0.4~\Delta_0$.
    }
    \label{fig:Cq_calibration}
\end{figure}

In this subsection, we calculate the quantum capacitance of parity states in double quantum dots.
The Hamiltonian for the double quantum dots is
\begin{align}
\hat{H}_{DQD} = \varepsilon_1 \hat{c}\dg_1 \hat{c}_1 + \varepsilon_2 \hat{c}\dg_2 \hat{c}_2 + t (\hat{c}\dg_2 \hat{c}_1 + \hat{c}\dg_1 \hat{c}_2)
 + \Delta ( \hat{c}\dg_1\hat{c}\dg_2 +  \hat{c}_2 \hat{c}_1 ),
\end{align}
where $\varepsilon_{1,2}$ are the onsite energies of the two dots, and $t$ and $\Delta$ are the normal and superconducting couplings between them.
To calculate the ground-state energies and the quantum capacitances (the second-order derivative of the ground-state energies), we use the occupation number basis $| n_1, n_2 \rangle = (\hat{c}\dg_1)^{n_1} (\hat{c}\dg_2)^{n_2} | 0 0 \rangle$ with $| 00 \rangle$ being the vacuum state.
Under this basis, the Hamiltonian can be decomposed into even- and odd-parity subspaces as below
\begin{align}
\hat{H}_{\rm{even}} =
\begin{pmatrix}
| 00 \rangle \\
| 11 \rangle
\end{pmatrix} \tp
\begin{pmatrix}
0 & \Delta \\
\Delta & \varepsilon_1 + \varepsilon_2
\end{pmatrix}
\begin{pmatrix}
\langle 00 | \\
\langle 11 |
\end{pmatrix}, \quad
\hat{H}_{\rm{odd}} =
\begin{pmatrix}
| 10 \rangle \\
| 01 \rangle
\end{pmatrix} \tp
\begin{pmatrix}
\varepsilon_1 & t \\
t & \varepsilon_2
\end{pmatrix}
\begin{pmatrix}
\langle 10 | \\
\langle 01 |
\end{pmatrix}.
\end{align}
After having the ground-state energies in each parity subspace, we take its second-order derivative with respect to gate voltage to obtain the value of quantum capacitance as follows
\begin{align}
    C_q = - \partial^2 E / \partial V^2_g,
\end{align}
where 
\begin{align}
    & \delta \varepsilon_1 =\alpha_1 \cdot \delta V_g =  \alpha \cos \theta \cdot \delta V_g \nn
& \delta \varepsilon_2 =\alpha_2 \cdot \delta V_g =  \alpha \sin \theta \cdot \delta V_g.
\end{align}
In the calculation, we particularly consider the calibration errors in the quantum capacitance measurement. 
That is, the parameter set $(E_L, E_R, t, \Delta)$ can be off the sweet spot to characterize the possible calibration errors in a realistic experiment. 
We consider two specific scenarios.
The first is errors in $E_L $ and $E_R$, while the second is in $t$ and $\Delta$.
We further assume that the errors in the four parameters are independent of each other.
For each parameter, its possible error is within a range of $ [-\delta, \delta]$ with uniform distribution. We take $\delta /\Delta_0=0, 0.2, 0.4$, where $\Delta_0$ is the strength of coupling without calibration errors.
The results of the numerical calculations are shown in Fig.~\ref{fig:Cq_calibration}.
In the absence of errors, the numerical calculations agree with the analytic results in Fig.~4.
In the presence of errors, it shows that quantum capacitance measurement is very robust against calibration errors in the onsite energies $E_i$, but more sensitive with the errors in the couplings. 
This behavior is consistent with the calculations performed for the fusion protocols in the main text.

\end{document}